\documentclass[twoside,12pt]{article}
\newcommand{\be}{\begin{equation}}
\newcommand{\ee}{\end{equation}}
\date{ }
\begin{document}

\setcounter{equation}{0}
\setcounter{section}{0}
\title{ \bf Intense laser interacting with a two level atom: WKB
expressions for
dipole transitions and  population inversion}
\author {{\bf Juan D. Lejarreta}\\ {\small \bf Escuela T\'ecnica Superior
de Ingenier\'{\i}a
Industrial}.  {\small \bf Universidad de Salamanca}\\  {\small \bf 37700.
B\' ejar. Spain}\\$$ $$\\
{\bf Jose M. Cerver\'o} \\ {\small \bf F\'{\i}sica Te\'orica}. {\small \bf
Facultad de Ciencias}.
{\small \bf Universidad de Salamanca}\\ {\small \bf 37008. Salamanca. Spain}}
\maketitle
\begin{abstract} In a previous paper, we have already considered the system
composed by a two
level atom interacting with a coherent external electromagnetic field. No
application whatsoever
has been made of the rotating wave approximation. Being specially
interested in the
problem of higher harmonic generations for the case of very intense laser
fields,  we
have developed in this letter a much more efficient  way to obtain these
solutions as
well as to carry out some calculations in a range in which the  parameters
take extreme values.
Also the formalism allows us now to provide analytic expressions in the WKB
regime for the electric dipole moment and the population inversion. The
spectrum can be decomposed in
periodic and non-periodic contributions. Only the latter depends upon the
Floquet exponent and can be responsible
of the main complexities of the observed Rabi revivals and the hyper-Raman
shift.

\end{abstract}
\vskip 0.4cm
$\qquad\qquad\qquad${\bf PACS Numbers: 03.65.Ca,  03.65.Fd  and  42.50.Hz}
\vskip 0.3 true in
\newpage

\noindent In a previous paper \cite{ICER-LEJ10} we have explored the
appearance of higher harmonics as an effect arising from the monochromatic
electric field of very
intense laser interacting with a two level atom. An adequate theoretical
description of the
interaction of the atom with the strong external field allow us, using  the
results presented in
\cite{ICER-LEJ6}, to build up the time evolution operator of the system and
exactly calculate
the instantaneous atomic dipole moment and the population inversion. We
obtain the Fourier transform
of these quantities and identifying the present frequencies and its
amplitudes we can explain the
composition of the spectrum of the outgoing radiation. In the present
letter an entirely new viewpoint
will be adopted sheding some light not only on the theoretical aspects
already discussed but also
enabling us to compute the correspondent physical quantities -either exactly or
approximate- in a way much more accessible to experimental verification. In
order to follow a self-contained
approach we shall be presenting first  a very brief account of
\cite{ICER-LEJ10} .

Let us consider the  physical system containing a {\bf two-level atom} and
its electric dipole
interaction with the coherent field of a laser \cite{ICER-LEJ10}. Let $|a>$
and $|b>$ be the atomic
eigenstates with energies are 0 and $\hbar\omega_0$ respectively. The
external monochromatic field
frequency is $\omega$ and  $\Omega_0$ is the $Rabi$ $frequency$
corresponding to an interaction
energy $\hbar \Omega_0$. In terms of the operators:
\begin{eqnarray}
J_0={1\over 2}\{|b><b|-|a><a|\}\qquad \qquad J_-=|a><b|\qquad \qquad
J_+=|b><a|
\end{eqnarray}
the system can be described by the well known Hamiltonian
\begin{eqnarray}
H\left(t\right)=\hbar \omega_0\left[J_0  + {1\over2} \right]+\hbar \Omega_0
\left[J_++J_-\right]\cos\omega t
\end{eqnarray}
Here $H(t)$ is a hermitian element of the triparametric $SU\left(2\right)$
Lie-Algebra
and the physical states of the atom can be obtained by acting on the the
eigenstates of
$J_0$ with the time evolution operator (See refs.
\cite{ICER-LEJ10}-\cite{ICER-LEJ6}) :
\begin{eqnarray}
U\left(t\right) = \exp\{-{i\omega_0\over 2}t\}\exp\{\eta (t) J_+\}
\exp\{\gamma (t)J_0\}
\exp\{- \eta^* (t) J_-\}\exp\{ih\left(t\right)J_0\}
\end{eqnarray}
where $\gamma (t) = Ln(1 + \mid\eta (t)\mid ^2) $ and:
\begin{eqnarray}
h\left(t\right)&=&-\omega_0t+2\Omega_0\int_0^t \cos \left(\omega
s\right)Re\left[\eta\left(s\right)\right]\,ds\\
\eta (t)&=&{\omega_0 q(t)+2i \dot q(t)\over \omega_0 q(t)-2i \dot q(t)}
\end{eqnarray}
The operator  $U(t)$ satisfies the Schr\"{o}dinger equation if the complex
function $q(t)$
is a solution of the following {\bf second order ordinary linear
differential equation} (ODE).
\begin{eqnarray}
\ddot q(t)-2 i \Omega_0 \cos (\omega t ) \dot q(t) + {\omega_0^2\over 4}
q(t) = 0
\qquad ; \quad q\left(0\right) =1\qquad ; \quad  \dot
q\left(0\right)={i\omega_0\over 2}
\end{eqnarray}
The main result that we would like to point out here is that
it is in fact possible to condense the full dynamics of the physical
problem in this
linear differential equation \cite{ICER-LEJ10}. Its solution determines all
the mathematical quantities
that will be considered of interest for the physical discussion of the
laser-atom interaction. For instance,
the rescaled Dipole Moment $D(t)$ and the Population Inversion $W(t)$ are
defined from the complex
solution $q(t) = r(t) \exp\{i\phi (t)\}$ of (6) as:
\begin{eqnarray}
D(t) &=& |q(t)|^2-1 = r^2(t)-1 \qquad\qquad\qquad W(t) = -{2\over \omega_0}
r^2(t)\dot\phi(t)
\end{eqnarray}
Using (6) and (7) we can easily check that the
following system of differential equations holds for $D(t)$ and $W(t)$:
\begin{eqnarray}
\ddot D(t)&+&\omega_0^2D(t)=2\omega_0 \Omega_0 \cos (\omega t)
W(t)\qquad;\qquad D(0) = 0\quad;\quad\dot D(0) = 0
\end{eqnarray}
\begin{eqnarray}
\dot W(t)&=& -{2\Omega_0 \over \omega_0}\cos(\omega t)\dot D(t)
\qquad\quad;\qquad\qquad W(0) = -1
\end{eqnarray}
and after some manipulations the following fundamental invariant can easily
be constructed:
\begin{eqnarray}
 \dot D^2(t) +\omega_0^2 D^2(t)+\omega_0^2 W^2(t)=\omega_0^2
\end{eqnarray}
which guarantees a bound solution for any set of parameters $(\omega_0\neq 0)$.
The differential system (8)-(9) has already been discussed in the
literature (See Refs. \cite{ICOM},\cite{ISUN} and
\cite{IKUL}). We have shown in \cite{ICER-LEJ10} the relationship of this
differential system with the ODE (6),
which has been proven to be crucial for establishing the non trivial
physical features to be discussed below.
In particular  two important properties of this differential system must be
emphasized: The existence of the first
integral  (10) and the direct relationship of $W(t)$ and $D(t)$  with a
complex function  $q(t) = r(t)\exp\{i\phi
(t)\}$ which turns out to be a solution of an ordinary linear differential
equation with periodic coefficients.

The description of the physical system in terms of $q(t)$ instead of the
dipole moment $D(t)$ and
the population inversion $W(t)$  is extremely advantageous as it allows us
to express directly
physical quantities such as the atomic radiation emitted by the excited
atom, its spectral
composition, the phases as well as the amplitudes of each component  and
the correspondent
relationship with the relevant parameters of the system (i.e. the atomic
frequency transition $\omega_0$,
the laser frequency $\omega$ and the Rabi frequency $\Omega_0$). At this
point we should mention the pioneer papers
by Shirley \cite{ISHI}, Zel'dovich \cite{IZEL}, Cohen-Tannoudji and Haroche
\cite{ICOH} and Eberly and coworkers
\cite{IEBE} which can be considered the first serious attempts to solve
this problem rigurously. There are
however many aspects which have been either overlooked or treated in a
different manner by these
authors. The results hereby presented allows us to analize in a much more
efficient way other interesting effects.
Among these we would like to mention the effect of the initial atomic state
as well as that of the initial phase
of the laser field with or without modulation terms. In order to achieve
all these goals it seems essential to us a
throughout analysis of the properties of $q(t)$ and hence those of $D(t)$ y
$W(t)$ \cite{ISUN}. To this end we shall
construct the Fourier spectrum of the relevant physical quantities. We
can also present several approximate formulae for the phase and the amplitude
for all present modes of the atomic emission spectra. Rescaling the time
variable in the form $x = \omega t$ and using $primes$ to denote
derivatives with
respect to $x$, the linear equation (6) reads:
\begin{eqnarray}
q''(x)-2i\gamma \cos xq'(x)+{\epsilon^2\over 4}q(x)=0\qquad \qquad q(0)=1\qquad
\qquad q'(0)=i{\epsilon\over 2}
\end{eqnarray}
where the two new dimensionless parameters $\gamma = {\Omega_0\over
\omega}$ and $\epsilon = {\omega_0\over\omega}$ are the ratio between the
interaction and foton laser and the  atomic and foton laser energies
respectively.

The ODE (11) can be
solved by applying Frobenius Theory \cite{ICER-LEJ10}. Let us define two
independent functions $u(x)$ and $v(x)$
satisfying $ u(0)=1, u'(0)=0, v(0)=0, v'(0)=1$. The Taylor coefficients are
given in \cite{ICER-LEJ10}. One can show
that:
\begin{itemize}
\item The general solution of (11) is a linear superposition of $u(x)$ and
$v(x)$. In particular, we set henceforth
$q(x) = u(x)+i{\epsilon\over 2}v(x)$
\item The equation posseses a {\bf first integral} $|q'|^2 +
{\epsilon^2\over 4}|q|^2 = C_o$ which allows us to set some bounds for the
solutions $u(x)$ and
$v(x)$ and yields as a consequence of the {\bf first integral} (11) the
absolute bound:
\begin{eqnarray}
|u^2(x)| +{\epsilon^2\over 4} |v^2(x)| =1
\end{eqnarray}
\item The set of functions $u(x) $ and $v(x)$ also satisfy:
\begin{eqnarray} u'(x) =  - {\epsilon^2\over 4}e^{2i\gamma \sin x}v^*(x)
\qquad\qquad\qquad\qquad
v'(x) =  e^{2i\gamma \sin x}u^*(x)
\end{eqnarray}
These relationships clearly show that $u(x)$ is $2\pi$-periodic
($2\pi$-antiperiodic) if and only if
$v(x)$ exhibits also the same properties. Thus, for any set of parameters
($\epsilon\neq 0$) for
which a solution of (11) exists with the property of being $2\pi$-periodic
($2\pi$-antiperiodic)
we can claim that all solutions will also be $2\pi$-periodic
($2\pi$-antiperiodic).
Notice also that $u(x)$ and $v(x)$ have the following symmetry properties:
\begin{eqnarray}
u(x+\pi)  =  u(\pi) u^*(x)-{\epsilon^2\over 4} v^*(\pi) v^*(x)
\end{eqnarray}
\begin{eqnarray}
v(x+\pi) =  v(\pi) u^*(x)+  u^*(\pi) v^*(x)
\end{eqnarray}
Also $u(-x) = u^*(x)$ and $v(-x) = - v^*(x)$. Clearly these properties can
be used to know the values of the
functions $u(x)$ and $v(x)$ for all $x$ if they are known just on the
$[0,\pi]$ interval. This interval may be
reduced in fact to $[0,{\pi\over 2}]$. In particular, ${1\over
2}(u(2\pi)+v'(2\pi)) = Re[u(2\pi)]= 1-2(\epsilon
Re[u({\pi\over 2}) v^*({\pi\over 2})])^2$.
\item The set of functions $D(x) $ and $W(x)$ may be expresed as:
\begin{eqnarray}
 D(x)=\epsilon Im(u(x) v^*(x)) \quad\quad\quad\quad  W(x)=- Re[e^{-2 i
\gamma sinx}(u^2(x)+
{\epsilon^2\over 4} v^2(x))]
\end{eqnarray}
\item As the equation (11) has periodic coefficients a Floquet analysis can
be performed.
According to this theorem any solution can be expressed in the form:
\begin{eqnarray} q(x) = A e^{i \nu x}F_+(x)+B e^{-i \nu x}F_-(x)
\end{eqnarray}
where $F_+(x)$ y $F_-(x)$ are $2 \pi$-periodic functions. Using the above
properties we end up with an explicit
expression for the Floquet exponent, namely:
\begin{eqnarray}
\nu =  {1\over 2\pi}\arccos [Re[u(2\pi)]]= {1\over \pi}\arcsin [\epsilon
Re[|u({\pi\over 2})
v^*({\pi\over 2})|]]\qquad\qquad  0\leq  \nu  \leq {1\over 2}
\end{eqnarray}
Its physical significance lies on the fact that it actually determines the
spectral decomposition
of the emitted atomic radiation. The two extreme cases correspond to
solutions $2\pi$-periodic
($\nu$ = 0) or $2\pi$-antiperiodic ($\nu = {1\over 2}$).

\end{itemize}
The Fourier spectrum of $u(x)$ and $v(x)$ takes the following generic form:
\begin{eqnarray} f(x) = A\sum_{j=-\infty}^{j=\infty}F_j e^{i(j+\nu )
x}+B\sum_{j=-\infty}^{j=\infty}(-1)^j F_{-j} e^{i(j-\nu ) x}
\end{eqnarray}
where the {\bf
Floquet exponent} $\nu$ and the {\bf Fourier coefficients} $F_j$ can easily
be found simultaneously
with the help of the following recurrence relation:
\begin{eqnarray} {1\over \gamma}  \{{(p + \nu )^2-{\epsilon^2\over 4}}\}
F_p  &=&(p + \nu +1)
F_{p+1}+(p + \nu -1)  F_{p-1}
\end{eqnarray}
which can be solved through a numerical method using {\bf continued fractions}
\cite{ICER-LEJ10}. The instantaneous dipole moment of the atom can finally
be expressed as:
\begin{eqnarray} D(x)=\epsilon Im(u(x) v^*(x)) = \Delta_1(x) +\Delta_2(x)
\end{eqnarray}
This expression is in fact the sum of two different contributions. The
$\Delta_1(x)$
term is a superposition of odd harmonics of the laser frequency. The
$\Delta_2(x)$ term is a superposition of even
harmonics shifted up and down an amount given by $2\nu$.
\begin{eqnarray}
\Delta_1(x) &=& \sum_{j=0}^\infty D_j \cos [(2 j+1)x] \\
\Delta_2(x) &=& \sum_{j=0}^\infty D^+_j \cos [2(j+\nu) x]+\sum_{j=1}^\infty
D_{j}^- \cos [2(j-\nu)
x]\end{eqnarray}
The correspondent Fourier amplitudes are:
\begin{eqnarray}
 D_j&=&2{M_e^2-M_o^2\over (M_e^2+M_o^2)^2}\sum_{r=-\infty}^\infty F_r
F_{r+2 j+1}\qquad\qquad\qquad\qquad M_e
=\sum_{r=-\infty}^\infty F_{2 r}\\
 D_j^\pm&=&-2{M_e M_o \over (M_e^2+M_o^2)^2}\sum_{r=-\infty}^\infty (-1)^r
F_{-r}
F_{r\pm 2j}\qquad\qquad\quad M_o =\sum_{r=-\infty}^\infty F_{2 r+1}
\end{eqnarray}
The Fourier spectrum of $D(x)$ dictates the composition of the emitted atomic
radiation in interaction with a laser. A qualitative picture shows the
appearance of a triplet
centered in the odd harmonics with frequencies $(2s+1 \pm\delta)\omega$.
The amount of the shift
$\delta$ is given by the Floquet exponent (18) in the form $\delta =1-2\nu $
and coincides with the generalized Rabi frequency. The non-periodic
contribution
$\Delta_2(x)$ of the instantaneous atomic dipole moment, which can be
interpreted as a superposition of odd harmonics
shifted up and down an amount $\delta$, is the origin of the hyper-Raman
peaks of the triplet.
Likewise,  the population inversion can also be cast in the form:
\begin{eqnarray} W(x)=-1-2 {\gamma\over\epsilon} \int_0^x cos z D'(z) dz =
\Pi_1(x)+\Pi_2(x)
\end{eqnarray}
as the sum of two contributions $\Pi_1(x)$ (a superposition
of even harmonics of the laser frequency)  and $\Pi_2(x)$ which is a
superposition of odd harmonics
shifted up and down an amount given by $2\nu$:
\begin{eqnarray}
\Pi_1(x)&=& \sum_{j=0}^\infty W_j \cos [ 2 j x]\\
\Pi_2(x)&=&\sum_{j=1}^\infty W_{j}^+ \cos [(2 j +1+2 \nu )x]+ \sum_{j=0}^\infty
W^-_j
\cos [ (2 j +1 - 2\nu)x]
\end{eqnarray}
with Fourier amplitudes given by:
\begin{eqnarray}
W_0 =  - (1 + W_{0}^-)-\sum_{j=1}^\infty (W_{j} +W_{j}^++W_{j}^-) \qquad\quad
\end{eqnarray}
\begin{eqnarray}
W_j = - {\gamma\over\epsilon} (D_j+D_{j-1}+{D_j-D_{j-1}\over 2 j})\qquad
;\quad   j\geq 1
\end{eqnarray}
\begin{eqnarray}
W_{0}^- = - {\gamma\over\epsilon} (D_{0}^++D_{1}^-+{D_{1}^--D_{0}^+\over
1-2\nu}) \qquad\qquad\qquad\quad
\end{eqnarray}
\begin{eqnarray}
W_{j}^\pm = - {\gamma\over\epsilon}
(D_{j}^\pm+D_{j+1}^\pm+{D_{j+1}^\pm-D_{j}^\pm\over 2 j+1\pm2\nu})
\qquad ; \quad  j\geq 1
\end{eqnarray}
Aside from the numerical procedure provided by (18), (24)-(25) and
(29)-(32), a complete analytical reconstruction
of the spectrum can be carried out starting with an input provided by the
$D(x)$ and $W(x)$ functions. If one can find analytic expressions for the
functions $u(x)$ and $v(x)$, the values for $D(x)$ and
$W(x)$ are inmediately known  from (16) and the amplitudes $D_j$,
$D^\pm_j$, $W_j$, and $W^\pm_j$ can easily be
found as:
\begin{eqnarray}
D_j&=& {4\over\pi}\int_0^{\pi\over 2}\Delta_1(x) \cos [(2j+1)x]dx
\qquad\qquad\qquad\qquad\qquad\\
\nonumber D_j^\pm &=& {2\over\pi}\int_0^{\pi\over 2}\Delta_2(x)\cos
[(2(j\pm\nu)x]dx
\pm\qquad\qquad\qquad\qquad\\
&\pm &{1\over\pi\sin(2\pi\nu)}\int_0^{\pi\over 2}\{\Delta_2(x-\pi)-\Delta_2(x+\pi)\}\sin [(2 (j\pm\nu)x]dx\quad\\
W_j &=& {2m_j\over\pi}\int_0^{\pi\over 2}\Pi_1(x) \cos [ 2j x]dx
\qquad\qquad\qquad\qquad\qquad\\
\nonumber W_j^\pm &=& {2\over\pi}\int_0^{\pi\over 2}\Pi_2(x)\cos [(2j+1\pm
2\nu)x]dx \mp \qquad\qquad\qquad\\
&\mp & {1\over\pi\sin (2\pi\nu)}\int_0^{\pi\over
2}\{\Pi_2(x-\pi)-\Pi_2(x+\pi)\}\sin [(2j+1\pm 2\nu)x]dx
\end{eqnarray}
Where $m_0$ =1 and $m_j$ =2 $(j\geq 1)$. These expressions are obviously
exact as long as so were the knowledge of
the Floquet exponent $\nu$ and  the functions $u(x)$ and $v(x)$.

As a concrete example of the above discussion,
we shall be considering the case in which the parameters of the physical
system verify $\gamma
\approx
\epsilon >> 1$. This condition holds for values of the laser frequency
smaller than the values of the resonance frequency
and the coupling constant
$\alpha = {\gamma
\over\epsilon}$ is of the order of one. In this way we are considering
subresonant systems
$\omega << \omega_o\approx \Omega_o$ with interaction and transition
energies of similar values. Since in other papers
\cite{ISUN},\cite{IKUL} a numerical study of this range of parameters has
been made we have chosen the same range for
the sake of comparison. This case seems also quite adequate for using the
$WKB$ method for high values of
the parameter $\epsilon$. In regard to the equations (11)-(13) this means
that we are looking for a solution of in the
form  $q(x) = exp\{i\gamma \sin x + \epsilon\sum_{k=0}^\infty z_k(x)
\epsilon^{-k}\}$, where the $ z_k(x)$ verify:
\begin{eqnarray}
z'^2_0(x) &=& - {1\over 4} (1+4\alpha^2\cos^2x)
\end{eqnarray}
\begin{eqnarray}
z'_k(x) &=&-{1\over 2z'_0(x)}
\{z''_{k-1}(x)+\sum_{j=1}^{k-1}z'_{j}(x)z'_{k-j}(x)\}\qquad\qquad k\geq 1
\end{eqnarray}
After calculating the sequence for higher orders of approximation, $u(x)$
and $v(x)$ can be expressed as:
\begin{eqnarray}
u(x)  &\approx&
{\sqrt\lambda e^{i\gamma\sin x}\over 2 (1+4\alpha^2\cos^2x)^{1\over
4}} \{\ (1-{2\alpha\over\lambda}) f_1(\alpha,\epsilon ,x) +
(1+{2\alpha\over\lambda}) f_2(\alpha,\epsilon, x)\}
\end{eqnarray}
\begin{eqnarray}
v(x)  &\approx& {ie^{i\gamma\sin x}\over \epsilon \sqrt \lambda}
{\{f_2(\alpha ,\epsilon ,x)
-f_1(\alpha ,\epsilon, x)\}\over (1+4\alpha^2\cos^2x)^{1\over 4}}
\end{eqnarray}
\begin{eqnarray}
f_1(\alpha ,\epsilon ,x)  &\approx&  [(\lambda+2\alpha)(\sqrt{1+4\alpha^2\cos^2
x}-2\alpha\cos x)]^{1\over 2}\Phi (\alpha ,\epsilon ,x)
\end{eqnarray}
\begin{eqnarray}
f_2(\alpha ,\epsilon ,x)  &\approx&
[(\lambda-2\alpha)(\sqrt{1+4\alpha^2\cos^2 x}+2\alpha \cos x)]^{1\over 2}
\Phi^* (\alpha ,\epsilon ,x)
\end{eqnarray}
\begin{eqnarray}
\Phi (\alpha ,\epsilon ,x)  &\approx& \exp
\{i({\epsilon\lambda E[x,k^2]\over2}+{(1+8\alpha^2)E[x,k^2]-F[x,k^2]\over
12\epsilon\lambda})\}
\end{eqnarray}
where $\lambda = \sqrt {1+4\alpha^2}$, $k^2 = {4\alpha^2\over
1+4\alpha^2}$ and $F[x,k^2]$ and $E[x,k^2]$ are the
incomplete elliptic integrals of the first and second kind respectively.
>From these expressions one can derive the values
of the Floquet exponent as well as the values of the instantaneous dipole
moment. These values take
the form:
\begin{eqnarray}
\Omega (\epsilon ,\alpha) &=& {1\over\pi}(\epsilon\lambda E[k^2]
+{(1+8\alpha^2)E[k^2]- K[k^2]\over 6\epsilon \lambda })\\
\nu (\epsilon ,\alpha) &\approx&  \arcsin [|\sin \Omega
(\epsilon,\alpha)|]\qquad\qquad\qquad 0\leq\nu\leq {1\over 2}
\end{eqnarray}
For sufficiently large values of $\epsilon$ and practically {\bf any value}
of $\alpha$
$(\epsilon >>\alpha )$ the Floquet exponent $\nu $ is a $2 \epsilon_o $ -
periodic
function of
$\epsilon $ of the form:
\begin{eqnarray}
\nu&\approx&{\epsilon\over 2\epsilon_0}\qquad\qquad\qquad if \qquad\qquad 0\leq
\epsilon\leq\epsilon_0\\\
\nu &\approx&1- {\epsilon\over 2\epsilon_0}\qquad\qquad if \qquad\qquad
\epsilon_0\leq
\epsilon\leq 2\epsilon_0
\end{eqnarray} where
$\epsilon_0 =\pi(2\sqrt {1+4\alpha^2} E[{4\alpha^2\over 1+4\alpha^2}])^{-1}$
and $\nu $ repeats itself every time that $\epsilon $ increases in
2$\epsilon_0$. For
a given value of the coupling these oscillations do not die away and keep
going even if the
laser frequency decreases. In particular when $\alpha =1$ (the energy of
interaction equals the energy
of the transition) the value of  $\epsilon_0$ is approximately {\bf 0.596086}.
Notice the predictive power of (44) and (45). One can use it for selecting
values of $\epsilon$ and $\alpha$
with a given Floquet exponent $\nu$. Thus the spectral composition of the
dipole moment and the population inversion
can be selected at will with a judicious choice of the parameters. Finally
one can easily conclude that:
\begin{eqnarray}
\epsilon \approx {r\pi\over 2\sqrt{1+4\alpha^2} E[k^2]}\{1+\sqrt{1-{2
E[k^2]((1+8\alpha^2)E[k^2]-K[k^2])\over 3r^2\pi^2}}\}
\end{eqnarray}
If $r \approx n$, $\nu = 0$ and the functions $u(x)$ and $v(x)$ are $2
\pi$-periodic. Both the dipole moment
$D(x)$  and the population inversion $W(x)$ are composed by pure even and
odd harmonics of the laser.
If  $r\approx n +{1\over 2}$, $\nu = {1\over 2}$, and $u(x)$ and $v(x)$ are
$2\pi$-antiperiodic. The
spectrum of $D(x)$ contains just odd harmonics and that of $W(x)$ just even
harmonics. If $r\approx{1\over2}(n +{1\over
2})$, $\nu = {1\over 4}$, and the functions $u(x)$ and $v(x)$ are
$4\pi$-antiperiodic. The atomic dipole moment can be
constructed in terms of its following two components:
\begin{eqnarray}
\Delta_1(\alpha ,x) &=& -{2\alpha\over\lambda} {\cos
x\over\sqrt{1+4\alpha^2\cos^2x}}
\end{eqnarray}
\begin{eqnarray}
\Delta_2(\alpha ,\epsilon ,x) &=& {2\alpha\over\lambda} {\cos \{\epsilon\lambda
E[x,k^2]+{1\over
6\epsilon\lambda}\{{(1+8\alpha^2)E[x,k^2]-F[x,k^2]}\}\}\over\sqrt{1+4\alpha^2\cos^2x}}
\end{eqnarray}
Notice that $\Delta_1(\alpha ,x)$ is independent of $\epsilon$ and it is
a purely even 2$\pi$-periodic term under the periodic translation
Therefore, this term enjoys the adequated
symmetry to represent the periodic part of the dipole oscillation. One can
easily identify a term like this
as an oscillation at the laser frequency, with modulated amplitude and just
composed by odd harmonics
of this frequency.  The term $\Delta_2(\alpha ,\epsilon ,x)$ has quite
different properties being periodic just for those
values of the parameters for which the Floquet exponent $\nu$ takes
rational values. The values of the instantaneous
dipole moment hereby found can be used to find its respective spectral
components. The periodic part yields the odd
harmonics spectrum with amplitude and phase are given by:
\begin{eqnarray}
D_j=-{8\alpha\over\pi\sqrt{1+4\alpha^2}}\int_0^{\pi\over 2}{\cos x \cos [(2
j+1)
x]\over\sqrt{1+4\alpha^2\cos^2x}}\qquad\qquad j\geq 0
\end{eqnarray}
For sufficiently lower laser frequencies (sufficiently large $\epsilon$)
the odd
harmonic spectrum results quite independent of this frequency. The
different amplitudes depend just
on the strenght of the laser-atom interaction and  smoothly decrease for
higher order harmonics.
In particular the amplitude and phase of the first harmonic can explicitely
be given by:
\begin{eqnarray} D_0=-{2\over\pi\alpha}(E[{4\alpha^2\over
1+4\alpha^2}]-{1\over1+4\alpha^2}{K[{4\alpha^2\over 1+4\alpha^2}]})
\end{eqnarray} The second contribution gives rise to the hyper-Raman
spectrum. For any set of the
physical parameters of the system, either $\Omega +\nu$ or $\Omega -\nu$
must be an integer
number. One can can construct the correspondent amplitudes in the form:
\begin{eqnarray}
D_j^+&=&{4\alpha\over\pi\lambda}\int_0^{\pi\over 2}{
\cos [(\epsilon\lambda +{1+8\alpha^2\over 6\epsilon\lambda}) E[x,k^2]-
{1\over 6\epsilon\lambda}K[x,k^2]\pm
2(j+\nu)x]\over\sqrt{1+4\alpha^2\cos^2x}}\quad j\geq 0
\end{eqnarray}
\begin{eqnarray}
D_j^-&=&{4\alpha\over\pi\lambda}\int_0^{\pi\over 2}{
\cos [(\epsilon\lambda +{1+8\alpha^2\over 6\epsilon\lambda}) E[x,k^2]-
{1\over 6\epsilon\lambda}K[x,k^2]\mp 2(j-\nu)x]
\over\sqrt{1+4\alpha^2\cos^2x}}\quad j\geq 1
\end{eqnarray}
where the upper sign holds for $\Omega +\nu$ an integer and the lower sign
for the integer being $\Omega -\nu$.
A large number of oscillations substantially lowers the value of the integral, the correspondent values of the amplitudes
leads one to consider easily the r
egion of the spectrum of maximal amplitudes. Therefore for the first case,
the
amplitude of the harmonics with frequencies $2j+2\nu$ will be negligible
while the harmonics with
frequencies $2j-2\nu$ will have significant amplitudes which can even have
similar intensities to
that of the odd harmonics. These large amplitudes will be distributed over
a region with values of
$2j$ close to $\Omega+\nu$. This behaviour justifies the typical harmonic
generation spectrum found by experiment
showing an intensity distribution that first decreases rather steeply as
the harmonic order
increases (the odd spectrum) and then after remaining almost flat for a
number of harmonics finally decreases
steeply again in a "plateau" form (the non-integral spectrum) \cite{IPLA}.
All these facts are in fairly good
agreement with the cases discussed in \cite{ICOM} where the shift  $\delta$
is given by an elliptic integral as in
our case, but the correspondent  intensities are given in terms of Bessel
functions. Likewise, the population
inversion can explicitely be  characterized by using the two contributions:
\begin{eqnarray}
\Pi_1(\alpha ,x) &=& - {1\over\lambda \sqrt{1+4\alpha^2\cos^2x}}
\end{eqnarray}
\begin{eqnarray}
\Pi_2(\alpha ,\epsilon ,x) &=&- {4\alpha^2\over\lambda} {cos x\cos
[(\epsilon\lambda
+{1+8\alpha^2\over 6\epsilon\lambda})E[x,k^2]-{1 \over
6\epsilon\lambda}F[x,k^2]]\over\sqrt{1+4\alpha^2\cos^2x}}
\end{eqnarray}
and as in the previous case $\Pi_1(\alpha ,x)$ is independent of
$\epsilon$. It is a purely
$\pi$-periodic term with the adequate symmetry to represent the periodic
part of the
population inversion whose Fourier spectrum is just composed by even
harmonics of the laser
frequency with amplitudes:
\begin{eqnarray}
W_j&=&-{2 m_j\over\pi\sqrt{1+4\alpha^2}}\int_0^{\pi\over 2}{ \cos [2
jx]\over\sqrt{1+4\alpha^2\cos^2x}}\qquad\qquad
\end{eqnarray}
Where $m_0$ =1 and $m_j$ =2 $(j\geq 1)$.  The term $\Pi_2(\alpha ,\epsilon
,x)$ depends
on the two characteristic parameters of the system and yields the rest of
the spectrum with
amplitudes:
\begin{eqnarray}
W_j^+&=&-{8\alpha^2\over\pi\lambda}\int_0^{\pi\over 2}{
\cos x\cos [(\epsilon\lambda +{1+8\alpha^2\over 6\epsilon\lambda}) E[x,k^2]-
{1\over 6\epsilon\lambda}K[x,k^2]\pm (2j+1+
2\nu)x]\over\sqrt{1+4\alpha^2\cos^2x}}\quad ;j\geq 0 \quad\quad
\end{eqnarray}
\begin{eqnarray}
W_j^-&=&-{8\alpha^2\over\pi\lambda}\int_0^{\pi\over 2}{
\cos x\cos [(\epsilon\lambda +{1+8\alpha^2\over 6\epsilon\lambda}) E[x,k^2]-
{1\over 6\epsilon\lambda}K[x,k^2]\mp (2j+1-2\nu)x]
\over\sqrt{1+4\alpha^2\cos^2x}}\quad ;j\geq 1\quad\quad
\end{eqnarray}
with the same convention of signs as the one given above.
\newpage
\noindent {\bf Conclusions.} In a previous paper \cite{ICER-LEJ10}, the
present authors had already considered the system
composed by a two level atom  interacting with a coherent external
electromagnetic field by solving the main
equations in the Schr\" odinger picture without using the rotating wave
approximation. We were
specially interested in the spectral composition of the atomic dipole
moment and the population inversion.
These quantities are determined by means of two functions $u(x)$ and
$v(x)$ which are independent solutions of an ordinary differential equation
with periodic coefficients.
Floquet analysis applied to this equation yields the frequencies which are
present in the spectra
and allows us to identify two different contributions in $D(x)$ (the atomic
dipole moment) and $W(x)$
(the population inversion) which yield the different spectral components
through quadratures.
In this letter we have presented {\bf analytical expressions for the
frequencies
and amplitudes within a particular range of values of the physical
parameters}.
These expressions may be used to analyze
the influence of the different parameters in the spectra.
Our results are also clearly of experimental interest in assigning
odd and Floquet-shifted character to each observed harmonic. This gives rise to
the decreasing and flat parts of the observed spectrum.
\vskip 0.3cm
\noindent {\bf  Acknowledgments.} This research has been supported in part
by {\bf DGICYT} under contract
{\bf PB98-0262}.

\end {document}